\documentclass[conference]{IEEEtran}
\IEEEoverridecommandlockouts

\usepackage{amsmath,amssymb}
\usepackage{graphicx}
\usepackage{booktabs}
\usepackage{siunitx}
\usepackage{url}
\usepackage{cite}
\usepackage{subcaption}
\usepackage{algorithm}
\usepackage{algpseudocode}
\usepackage{comment}
\usepackage{soul}
\usepackage[nolist]{acronym}
\acrodef{3GPP}{3rd Generation Partnership Project}
\acrodef{5G}{fifth generation}
\acrodef{6G}{sixth generation}
\acrodef{BCGA}{block-coordinate gradient ascent}
\acrodef{BLASTER}{Bandwidth SpLit, User ASsociation, and PowEr ContRol}
\acrodef{BS}{base station}
\acrodef{CDF}{cumulative distribution function}
\acrodef{CL}{clutter loss}
\acrodef{Dec-Alg}{decentralized parallel execution algorithm}
\acrodef{FSO}{free-space optical}
\acrodef{GDC}{geometric disk cover}
\acrodef{GEO}{geostationary Earth orbit}
\acrodef{HAPS}{high-altitude platform station}
\acrodefplural{HAPS}[HAPS]{high-altitude platform stations}
\acrodef{LEO}{low Earth orbit}
\acrodef{LoS}{line-of-sight}
\acrodef{MBS}{macro base station}
\acrodef{MEC}{minimum enclosing circle}
\acrodef{MINLP}{mixed-integer non-linear programming}
\acrodef{NOMA}{non-orthogonal multiple access}
\acrodef{NT}{non-terrestrial}
\acrodef{NTN}{Non-terrestrial network}
\acrodef{O2I}{outdoor-to-indoor}
\acrodef{QoS}{quality of service}
\acrodef{RE}{resource element}
\acrodef{RSMA}{rate-splitting multiple access}
\acrodef{RSRP}{reference signal received power}
\acrodef{SCA}{successive convex approximation}
\acrodef{SE}{spectral efficiency}
\acrodef{SF}{shadow fading}
\acrodef{SINR}{signal-to-interference-plus-noise ratio}
\acrodef{SLT}{sum log-throughput}
\acrodef{TDM}{time-division multiplexing}
\acrodef{TN}{terrestrial network}
\acrodef{TR}{technical report}
\acrodef{UE}{user equipment}

\begin{document}

\title{HAPS-Enabled 6G TN-NTNs: Achieving Superior QoS-Energy Tradeoffs Compared with LEO}

\author{
\IEEEauthorblockN{
Azim Akhtarshenas\IEEEauthorrefmark{1}\IEEEauthorrefmark{2},
Henri Alam\IEEEauthorrefmark{3},
Matteo Bernab\`{e}\IEEEauthorrefmark{1},
Halim Yanikomeroglu \IEEEauthorrefmark{4}
and David L\'{o}pez-P\'{e}rez\IEEEauthorrefmark{1}
}
\IEEEauthorblockA{\IEEEauthorrefmark{1}iTEAM Research Institute, Universitat Polit\`{e}cnica de Val\`{e}ncia, Valencia, Spain\\
\IEEEauthorrefmark{2}Eindhoven University of Technology (TU/e), Eindhoven, The Netherlands\\
\IEEEauthorrefmark{3}Nokia Standards France, Nokia, France\\
\IEEEauthorrefmark{4}Carleton Non-Terrestrial Networks Lab, Carleton University, Ottawa, ON, Canada
}

\thanks{
This research is supported by the Generalitat Valenciana through the CIDEGENT PlaGenT, Grant CIDEXG/2022/17, Project iTENTE, and the action CNS2023-144333, financed by MCIN/AEI/10.13039/501100011033 and the European Union ``NextGenerationEU''/PRTR.
}
}

\newenvironment{bluenv}{\color{blue}}{}
\newenvironment{redenv}{\color[rgb]{1,0,0}}{}
\newenvironment{greenenv}{\color[rgb]{0,0.5,0}}{}
\newenvironment{brownenv}{\color[rgb]{0.65,0.16,0.16}}{}
\newenvironment{purplenv}{\color[rgb]{0.75,0,1.0}}{}
\renewcommand{\figurename}{Figure}

\maketitle




\begin{abstract}
Non-terrestrial networks (NTNs) can reduce terrestrial energy consumption by absorbing traffic from lightly loaded base stations (BSs), allowing them to be switched off. We study how the NTN platform shapes this quality of service (QoS)–energy trade-off, comparing one low Earth orbit (LEO) satellite at 600 km with 1, 4, and 9 high-altitude platform stations (HAPS) at 20 km. Using a traffic-aware framework that jointly optimizes user association, NTN bandwidth, and terrestrial BS activation over a 24-hour traffic profile, we show that 9 HAPS reduce terrestrial energy consumption by 34.1\%, versus 26.0\% for LEO. At the same QoS, 9 HAPS save 20\% more terrestrial energy than LEO while improving cell-edge throughput.
\end{abstract}



\section{Introduction}

\Acp{NTN} are emerging as a key component of \ac{6G} wireless architectures, 
complementing terrestrial networks to provide ubiquitous connectivity and enhanced network flexibility~\cite{svistunov2025bridging}.
Beyond extending coverage to remote and underserved regions, 
\acp{NTN} offer an opportunity to improve terrestrial network energy efficiency by dynamically offloading traffic and enabling underutilized \acp{BS} to be switched off~\cite{song2024high}.
This is particularly relevant in dense cellular deployments, 
where terrestrial infrastructure must remain active to guarantee continuous coverage, 
even during periods of low traffic demand.
\ac{LEO} satellite constellations have attracted significant attention for their ability to provide wide-area connectivity, high-capacity backhaul, and direct \ac{UE} access~\cite{nguyen2023two,nguyen2024joint}.
However, their high mobility introduces frequent handovers and complex resource-management challenges, 
while long propagation distances and atmospheric impairments can degrade communication performance, 
particularly at higher frequencies~\cite{madoery2024novel,li2025efficient}.
\ac{HAPS}, operating quasi-stationarily in the stratosphere, offer complementary advantages, 
including persistent regional coverage, lower propagation delays, and flexible deployment~\cite{svistunov2025bridging,cianca2005integrated,xing2021high}.
\ac{HAPS} have been investigated for satellite-terrestrial relaying~\cite{madoery2024novel,pace2004integrated} and interference coordination in integrated networks~\cite{ji2023joint}.
Recent research has further demonstrated the potential of \ac{HAPS} to enhance spectral and energy efficiency through advanced resource-management techniques.
For example, dynamic beamforming and \ac{NOMA} improve \ac{SE} and reduce outage probability~\cite{javed2024system}, 
while \ac{RSMA} enables improved resource allocation across integrated \ac{LEO}-\ac{HAPS}-terrestrial networks~\cite{tran2025joint}.
Energy-aware \ac{HAPS} operation,
including standby and deep-sleep mechanisms, 
has also been investigated to reduce overall network consumption~\cite{raj2024stochastic}.
Nevertheless, these approaches primarily focus on improving \ac{NTN} communication performance rather than exploiting its wide-area coverage to fundamentally optimize the operation of terrestrial infrastructure.
In this context,
the recently proposed Hypercell concept~\cite{song2024high} introduces a new architectural perspective in which a wide-area \ac{HAPS} coverage layer assumes part of the coverage responsibility traditionally assigned to terrestrial macro cells.
This enables both terrestrial coverage and capacity cells to be dynamically deactivated according to traffic demand, 
opening new opportunities for network energy savings.
Complementarily, Alam \textit{et al.}~\cite{alam2023throughput,alam2024optimizing} developed BLASTER, 
a traffic-aware radio resource management framework that jointly optimizes user association, bandwidth allocation, transmit power, and terrestrial \ac{MBS} activation in integrated \ac{TN}-\acp{NTN}.
By exploiting \ac{NTN}-assisted traffic offloading, 
BLASTER demonstrates significant energy savings while preserving user performance under time-varying traffic conditions.
However, the effectiveness of this approach strongly depends on the \ac{NTN} deployment configuration.
While \ac{LEO} satellites offer extensive coverage,
\ac{HAPS} provide shorter propagation distances and can be deployed in multiple coordinated platforms to increase spatial capacity.
Understanding how these architectural differences influence terrestrial \ac{MBS} shutdown, network throughput, and energy consumption is essential for determining the most effective operating points of future Hypercell-inspired \ac{TN}-\ac{NTN} architectures.
\subsection{Motivation and Contributions}
Motivated by the Hypercell concept~\cite{bernabe2026haps} and the BLASTER framework~\cite{alam2023throughput,alam2024optimizing}, 
this paper investigates the \ac{QoS}--energy consumption trade-off of integrated \ac{TN}-\acp{NTN} under different \ac{HAPS} deployment densities, using \ac{LEO} as a reference.
Our main contributions are listed as follows:
\begin{itemize}
    \item 
    We extend BLASTER to multi-\ac{HAPS} deployments comprising 1, 4, and 9 \ac{HAPS}, 
    and compare their performance against a \ac{LEO}-based architecture under realistic diurnal traffic variations.
    \item 
    We investigate how increasing the number of \ac{HAPS} affects \ac{UE} association, bandwidth allocation, network throughput, TN \ac{MBS} shutdown, and energy consumption.
    \item 
    We characterize the \ac{QoS}-energy consumption trade-off through mathematical analysis and system-level simulations, quantifying the additional terrestrial \ac{MBS} shutdown opportunities enabled by different \ac{NTN} configurations under comparable \ac{QoS} requirements.
\end{itemize}

\begin{table}[!t]
\vspace*{2pt}       
    \centering
    \caption{Normalized \ac{UE} density over a 24-hour period~\cite{alam2023throughput}.}
    \label{proportion_hour}
    \setlength{\tabcolsep}{3pt}   
    \begin{tabular}{cccccccc}
        \toprule
        \textbf{Hour} & \textbf{Density} &
        \textbf{Hour} & \textbf{Density} &
        \textbf{Hour} & \textbf{Density} &
        \textbf{Hour} & \textbf{Density} \\
        \midrule
        1  & 0.40 & 7  & 0.06 & 13 & 0.95 & 19 & 0.92 \\
        2  & 0.30 & 8  & 0.10 & 14 & 0.97 & 20 & 0.93 \\
        3  & 0.15 & 9  & 0.25 & 15 & 0.95 & 21 & 1.00 \\
        4  & 0.09 & 10 & 0.45 & 16 & 0.93 & 22 & 0.85 \\
        5  & 0.06 & 11 & 0.65 & 17 & 0.90 & 23 & 0.75 \\
        6  & 0.04 & 12 & 0.80 & 18 & 0.88 & 24 & 0.55 \\
        \bottomrule
    \end{tabular}
\end{table}

\section{System Model}
\label{sec:model}

We consider a heterogeneous terrestrial cellular network deployed over a geographical area $\mathcal{A}$ of size $A$ (km$^2$), 
partitioned into urban ($\mathcal{A}_{\mathrm{u}}$) and rural ($\mathcal{A}_{\mathrm{r}}$) regions,
such that $\mathcal{A}=\mathcal{A}_{\mathrm{u}}\cup\mathcal{A}_{\mathrm{r}}$ and $A=A_{\mathrm{u}}+A_{\mathrm{r}}$.
The set of active \acp{UE}, 
denoted by $\mathcal{U}(t)$, 
is spatially distributed according to regional population densities.
Their number varies following a normalized diurnal traffic profile, 
$|\mathcal{U}(t)|=K_{\mathrm{UE}}^{\max}\rho(t)$, where $\rho(t)\in(0,1]$ is given in Table~\ref{proportion_hour}, 
and $K_{\mathrm{UE}}^{\max}$ is the maximum number of active \acp{UE}.
A fraction $\eta_{\mathrm{in}}$ of \acp{UE} is located indoors, 
while the remaining $1-\eta_{\mathrm{in}}$ is outdoors.
Let $\mathcal{T}$ and $\mathcal{S}$ denote the sets of terrestrial and \ac{NT} (\ac{LEO}/\ac{HAPS}) \acp{MBS}, respectively, 
with cardinalities $T=|\mathcal{T}|$ and $S=|\mathcal{S}|$.
The complete set of \acp{MBS} is $\mathcal{B}=\mathcal{T}\cup\mathcal{S}=\{1,\dots,L\}$, 
with $L=T+S$.
The \ac{NT} tier complements terrestrial coverage, 
enabling traffic offloading and the shutdown of underutilized terrestrial \acp{MBS} while maintaining service continuity.
\begin{figure}[t]
    \centering
    \includegraphics[width=0.8\linewidth]{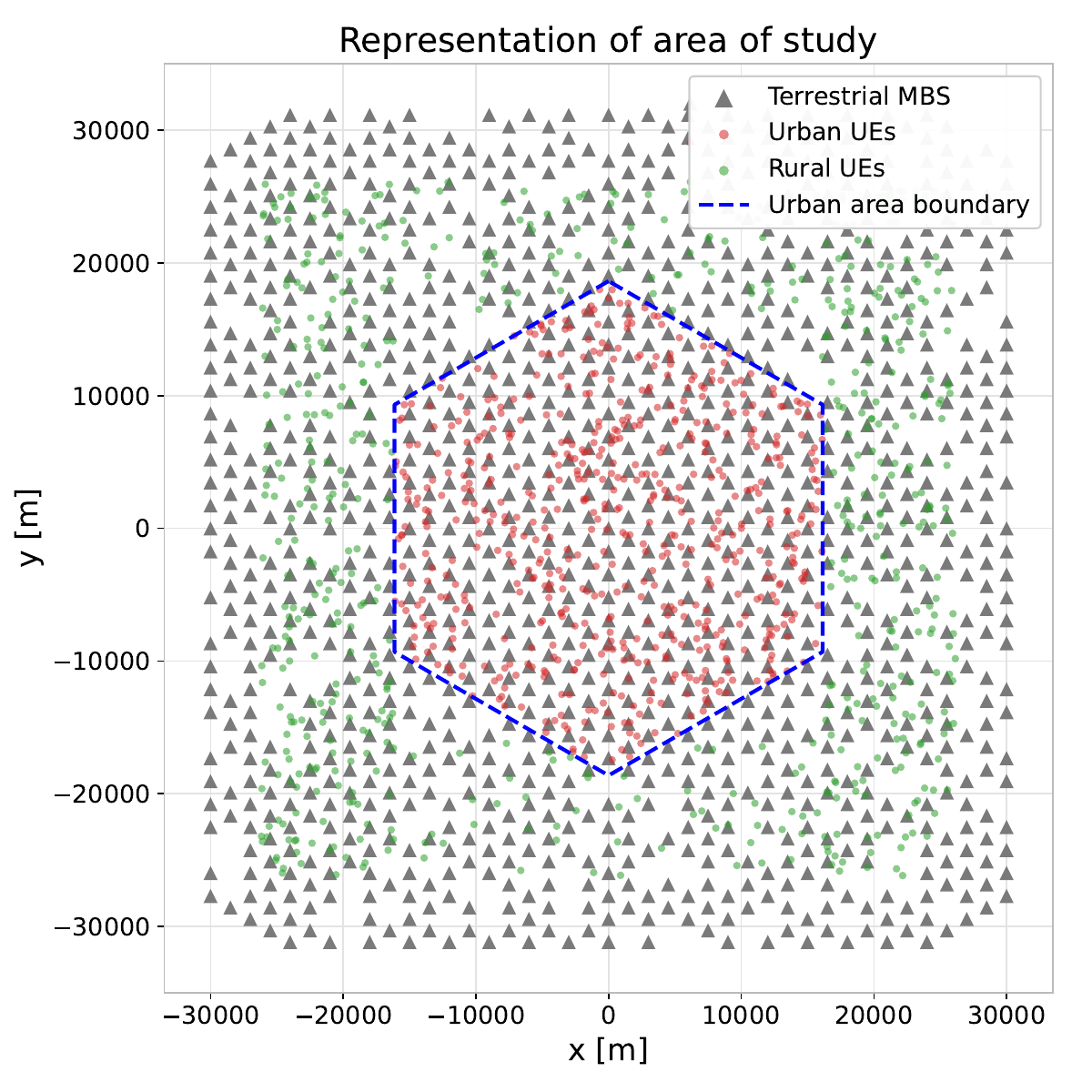}
    \caption{Study area with \acp{MBS} and urban/rural \acp{UE}.}
    \label{fig:AoS}
\end{figure}
Table~\ref{tab:params} summarizes the system parameters,
following the recommendations in~\cite{3gpp_tr38901_2022,3gpp_tr38811_2020,3gpp38821} as well as in~\cite{alam2024optimizing}.

\subsubsection{Channel Model}

We adopt the \ac{3GPP} terrestrial channel model in TR~38.901~\cite{3gpp_tr38901_2022} and its extension to satellite and \ac{HAPS} links in TR~38.811~\cite{3gpp_tr38811_2020}.

\textit{Terrestrial Link:}
Following the \ac{3GPP} channel model in~\cite{3gpp_tr38901_2022} and its formulation in~\cite{alam2024optimizing},
the large-scale channel gain between \ac{UE} $i$ and terrestrial \ac{MBS} $j$, in dB, is expressed as
\begin{equation}
    \beta_{ij}^{\mathrm{dB}} =
    G_j+G_i-PL_{ij}^{\rm b}-SF_{ij}
    -PL_{ij}^{\rm tw}-PL_{ij}^{\rm in}-N_{ij},
\end{equation}
where $G_j$ and $G_i$ denote the \ac{MBS} and \ac{UE} antenna gains, 
$PL_{ij}^{\rm b}$ is the basic path loss defined in Table~7.4.1-1 of~\cite{3gpp_tr38901_2022},
and $SF_{ij}\sim\mathcal{N}(0,\sigma_{SF}^2)$ denotes shadow fading.
The \ac{O2I} penetration loss comprises external-wall loss
$PL_{ij}^{\rm tw}$, indoor loss $PL_{ij}^{\rm in}$, and a random component $N_{ij}\sim\mathcal{N}(0,\sigma_p^2)$,
following Section~7.4.3 of~\cite{3gpp_tr38901_2022}.

\textit{NT Link:}
Following TR~38.811~\cite{3gpp_tr38811_2020},
the large-scale channel gain for a satellite or \ac{HAPS} link is
\begin{equation}
    \beta_{ij}^{\mathrm{dB}} =
    G_j+G_i-
    \left(PL_{ij}^{\rm b}+PL_{ij}^{\rm g} +PL_{ij}^{\rm s}+PL_{ij}^{\rm e}\right),
\end{equation}
where $PL_{ij}^{\rm b}$ comprises free-space path loss, shadow fading, and clutter loss,
while $PL_{ij}^{\rm g}$, $PL_{ij}^{\rm s}$, and $PL_{ij}^{\rm e}$ denote gaseous attenuation,
scintillation, and building entry loss, respectively.
All nodes employ omnidirectional antennas, with the corresponding antenna gains for terrestrial MBSs, the LEO satellite, HAPSs, and UEs specified in Table~\ref{tab:params}.

\subsubsection{SINR and Rate}

The downlink \ac{SINR} experienced by \ac{UE} $i$ associated with \ac{MBS} $j$ is
\begin{equation}
    \gamma_i =
    \frac{\beta_{ij}p_j}
    {\displaystyle\sum_{k\in\mathcal{I}_j}\beta_{ik}p_k+\sigma^2},
    \label{eq:sinr}
\end{equation}
where $\beta_{ij}=10^{\beta_{ij}^{\mathrm{dB}}/10}$ is the linear-scale channel gain,
$p_j$ is the transmit power per \ac{RE}, 
and $\sigma^2$ is the noise power per \ac{RE}.
Since terrestrial and \ac{NT} tiers use orthogonal spectrum portions, 
interference is restricted to the serving tier:
\begin{equation}
\mathcal{I}_j =
\begin{cases}
\mathcal{S}\setminus\{j\}, & j\in\mathcal{S},\\
\mathcal{T}\setminus\{j\}, & j\in\mathcal{T}.
\end{cases}
\end{equation}
Thus, multiple \ac{HAPS} reuse the same \ac{NT} spectrum and generate intra-tier interference.
Assuming equal bandwidth sharing among the $k_j$ \acp{UE} associated with \ac{MBS} $j$,
the achievable downlink rate is
\begin{equation}
    R_i=\frac{W_j}{k_j}\log_2(1+\gamma_i),
    \label{eq:throughput}
\end{equation}
where $W_j$ is the bandwidth available at the serving \ac{MBS}.

\subsubsection{Power Consumption Model}

We adopt the simplified \ac{MBS} power consumption model of~\cite{piovesan2022machine},
\begin{equation}
    Q_j(p_j)=P_0+p_j+\psi_j\lVert p_j\rVert_0,
    \label{eq:energy}
\end{equation}
where $P_0$ denotes the residual power consumption when the \ac{MBS} is deactivated, $\psi_j$ is the additional static consumption required for active operation, and $p_j$ represents the transmission-dependent power component.
The indicator $\lVert p_j\rVert_0$ equals one when $p_j>0$ and zero otherwise.
Thus, shutting down an \ac{MBS} eliminates its active-state static and transmission-dependent consumption, 
leaving only $P_0$.
This motivates \ac{NT}-assisted traffic offloading as a means of reducing \ac{TN} energy consumption.

\begin{table}[t]
\vspace*{4pt}        
    \centering
    \caption{System Parameters}
    \label{tab:params}
    
    \begin{tabular}{@{}
        p{4.47cm}
        p{1.45cm}
        p{1.85cm}
    @{}}
    \toprule
    \textbf{Parameter} & \textbf{Notation} & \textbf{Value} \\
    \midrule
    
    Area of study 
        & -- 
        & $\approx$\SI{2500}{km^2} \\
    
    Total bandwidth 
        & $W$ 
        & \SI{40}{MHz} \\
    
    Carrier frequency 
        & $f_c$ 
        & \SI{2}{GHz} \\
    
    Subcarrier spacing 
        & -- 
        & \SI{15}{kHz} \\
    
    Urban / rural inter-site distance \cite{alam2023throughput}
        & -- 
        & \SI{500}{m} / \SI{1732}{m} \\
    
    Urban / rural UE distribution
        & -- 
        & \SI{40}{\%} / \SI{60}{\%} \\
    
    Number of terrestrial MBS \cite{alam2023throughput}
        & $T$ 
        & 1776 \\
    
    Fraction of indoor UEs
        & -- 
        & \SI{80}{\%} \\
    
    Min. UE count (06:00) / Max. (21:00)
        & $N_{\mathrm{UE}}^{\min}/N_{\mathrm{UE}}^{\max}$
        & 400 / 10{,}000 \\
    
    LEO altitude
        & $h_{\mathrm{LEO}}$ 
        & \SI{600}{km} \\
    
    HAPS altitude
        & $h_{\mathrm{HAPS}}$ 
        & \SI{20}{km} \\
    
    Terrestrial max. Tx power/RE \cite{alam2023throughput}
        & $p_{j,\max}^{\mathrm{Ter}}$ 
        & \SI{17.7}{dBm} \\
    
    Satellite/HAPS max. Tx power/RE \cite{alam2023throughput}
        & $p_{j,\max}^{\mathrm{Sat/HAPS}}$ 
        & \SI{15.8}{dBm} \\
    
    
    
    Terrestrial antenna gain (fixed) \cite{alam2023throughput} & $G^{\mathrm{Ter}}_{\mathrm{TX}}$ & 14 dBi \\
LEO/HAPS antenna gain (fixed) \cite{alam2023throughput} & $G^{\mathrm{Sat/HAPS}}_{\mathrm{TX}}$ & 30 dBi \\
UE antenna gain (omni) \cite{alam2023throughput} & $G_{\mathrm{UE}}$ & 0 dBi \\
    
    Terrestria shadowing loss (SF) \cite{alam2023throughput}
        & $\mathrm{SF}_{\mathrm{Ter}}$ 
        & \SIrange{4}{8}{dB} \\
    
    Satellite/HAPS shadowing loss \cite{alam2023throughput}
        & $\mathrm{SF}_{\mathrm{Sat/HAPS}}$ 
        & \SIrange{0}{12}{dB} \\
    
    Noise power spectral density
        & $N_0$ 
        & \SI{-174}{dBm/Hz} \\
    
    Minimum RSRP for coverage
        & $\mathrm{RSRP}_{\min}$ 
        & \SI{-120}{dBm} \\
    
    Satellite energy consumption \cite{alam2023throughput}
        & $E_c$ 
        & \SI{500}{J} \\
    
    Baseline (idle) MBS power \cite{alam2023throughput}
        & $P_0$ 
        & \SI{25}{W} \\
    
    Static active-power term \cite{alam2023throughput}
        & $\psi_j$ 
        & \SI{50}{W} \\
    
    \bottomrule
    \end{tabular}
\end{table}

\section{Problem Formulation}
\label{sec:formulation}

We extend the LEO-based BLASTER framework~\cite{alam2024optimizing,alam2023throughput} to the proposed multi-\ac{HAPS} architecture.
Let $\mathbf{X}=[x_{ij}]\in\{0,1\}^{K\times L}$ denote the \ac{UE}-\ac{MBS} association matrix and $\mathbf{p}=[p_1,\ldots,p_L]^T\in\mathbb{R}_{+}^{L}$ the transmit power vector.
The fraction $\varepsilon\in[0,1]$ of the total bandwidth $W$ is allocated to the \ac{NT} tier, 
with $W_j=\varepsilon W$ for $j\in\mathcal{S}$ and $W_j=(1-\varepsilon)W$ for $j\in\mathcal{T}$,
ensuring orthogonal spectrum allocation between tiers.
The NT transmit powers are fixed to their maximum values,
i.e., $p_j=p_{j,\max}$ for $j\in\mathcal{S}$,
while only \ac{TN} transmit powers are optimized.
The joint optimization problem is expressed as
\cite{alam2024optimizing}
:
\begin{align}
\max_{\mathbf{X},\varepsilon,\mathbf{p}}\quad&
\sum_{i\in\mathcal{U}}\log(R_i)
-\lambda\sum_{j\in\mathcal{T}}Q_j(p_j)
\label{eq:obj}\\
\text{s.t.}\quad&
\sum_{j\in\mathcal{B}}x_{ij}=1,\quad
x_{ij}\in\{0,1\},\ \forall i,j, \nonumber\\
&
(\mathbf{X}\odot\boldsymbol{\beta})\mathbf{p}
\geq \mathrm{RSRP}_{\min}\mathbf{1}_K,\nonumber\\
&
0\leq p_j\leq p_{j,\max},\ \forall j\in\mathcal{B},
\quad 0\leq\varepsilon\leq1. \nonumber
\end{align}
Here, $\boldsymbol{\beta}\in\mathbb{R}^{K\times L}$ contains the linear-scale channel gains,
$\odot$ denotes element-wise multiplication,
$\mathbf{1}_K$ is the all-ones vector, and
$\lambda$ controls the trade-off between \ac{SLT} and terrestrial power consumption.
The logarithmic utility promotes proportional fairness.
To address the discontinuous activation-dependent power consumption in~\eqref{eq:energy},
BLASTER adopts a reweighted sparsity-promoting relaxation~\cite{alam2024optimizing}:
To handle the discontinuous activation term in~\eqref{eq:obj}, BLASTER adopts a reweighted sparsity-promoting relaxation~[16]:
\begin{equation}
\max_{\mathbf{X},\varepsilon,\mathbf{p}} \; \sum_{i\in\mathcal{U}} \log(R_i) - \lambda \sum_{j\in\mathcal{T}} \left( p_j + \phi\, w_j\, p_j^2 \right),
\label{eq:relaxed}
\end{equation}
subject to the constraints of~\eqref{eq:obj}.
The weights are iteratively updated as $w_j\leftarrow 1/(p_j+\delta)$,
where $\delta>0$ ensures numerical stability,
and $\phi$ controls the contribution of the reweighted quadratic penalty.
To adapt the throughput--energy trade-off to traffic variations,
we set $\lambda=\lambda_{\max}K_{\min}/K$,
where $K_{\min}$ and $K$ denote the minimum and current numbers of active UEs, respectively.
\begin{algorithm}[!b]
\caption{Block-Coordinate Gradient Ascent (BCGA)~\cite{alam2023throughput}}
    \label{alg:bcga}
    \begin{algorithmic}[1]
    
    \State Initialize $\mathbf{X}^{(0)}$ and $\mathbf{p}^{(0)}$
    \State Set $s \gets 0$
    \Repeat
        \State Update $\mathbf{X}^{(s+1)}$ using gradient projection,
        while enforcing
        $\tilde{\boldsymbol{\beta}}\cdot\mathbf{p}
        \geq \mathrm{RSRP}_{\min}\mathbf{1}_K$
        via Lagrangian duality
    
        \State Update bandwidth:
        $\varepsilon^{(s+1)} \gets K_S/K$
    
        \State Compute
        $\tilde{\mathbf{p}}^{(s)}
        \gets \mathbf{p}^{(s)}
        +\eta\nabla_{\mathbf{p}}f$
    
        \State Apply block soft-thresholding:
        \[
        \hat{\mathbf{p}}^{(s)}
        \gets
        \max\left\{
        1-\frac{t}{\|\tilde{\mathbf{p}}^{(s)}\|_2},0
        \right\}
        \tilde{\mathbf{p}}^{(s)},
        \qquad
        t=\lambda\eta\mathbf{w}^{T}\boldsymbol{\psi}
        \]
    
        \State Project $\hat{\mathbf{p}}^{(s)}$ onto
        $[\tau_j,p_{j,\max}]$, where
        \[
        \tau_j=
        \max_{i\in\mathcal{U}_j}
        \left(
        \frac{\mathrm{RSRP}_{\min}}{\beta_{ij}}
        \right)
        \]
        \State Set $p_j^{(s+1)} \gets 0$ for all $j \in \mathcal{T}$ with $k_j = 0$
    
        \State $s \gets s+1$
    
    \Until{convergence}
    \State \Return $\mathbf{X}^{\ast}$, $\varepsilon^{\ast}$, $\mathbf{p}^{\ast}$
    \end{algorithmic}
\end{algorithm}
BLASTER is solved independently for each hour of the day. A terrestrial MBS is deactivated when no UE is associated with it or when the reweighted sparsity penalty drives its transmit power to zero; deactivated MBSs may be reactivated in subsequent hours as traffic increases.
For fixed association $\mathbf{X}$ and transmit power $\mathbf{p}$, 
the optimal \ac{NT}-tier bandwidth fraction is $\varepsilon^\ast=K_S/K$,
where $K_S$ denotes the number of \acp{UE} associated with the \ac{NT} tier and $K$ the total number of \acp{UE}.
Thus, bandwidth allocation follows the fraction of \acp{UE} served by the \ac{NT} tier.
Problem~\eqref{eq:relaxed} is solved using \ac{BCGA},
as described in Algorithm~\ref{alg:bcga}.

\paragraph{Terrestrial per-\ac{UE} rate and MBS shutdown.}

Let $K_S$ and $K_T=K-K_S$ denote the numbers of \acp{UE} served by the \ac{NT} and terrestrial tiers, respectively,
and $M_{\mathrm{on}}$ the number of active terrestrial \acp{MBS}.
Assuming approximately uniform loading, 
each active \ac{MBS} serves $\bar{k}=K_T/M_{\mathrm{on}}$ \acp{UE}.
Using $\varepsilon^\ast=K_S/K$,
\eqref{eq:throughput} yields
\begin{equation}
    \bar{R}_{TN}\approx
    \frac{(1-\varepsilon^\ast)W}{\bar{k}}
    \mathbb{E}[\log_2(1+\gamma_i)]
    =
    \frac{WM_{\mathrm{on}}}{K}
    \mathbb{E}[\log_2(1+\gamma_i)].
    \label{eq:tn_rate}
\end{equation}
Initially, \ac{NT} offloading reduces terrestrial cell loads, 
enabling lightly loaded \acp{MBS} to be shut down.
Their remaining \acp{UE} are reassociated with active \acp{MBS}, 
potentially increasing their load and reducing the bandwidth available per \ac{UE}.
Furthermore, reassociation may increase the serving distance and path loss, 
reducing the desired received signal power.
Although shutting down \acp{MBS} also reduces interference,
the received signal degradation may dominate, 
resulting in lower mean \ac{SINR} and spectral efficiency.
Consequently, as additional offloading enables further shutdowns,
$\partial M_{\mathrm{on}}/\partial\varepsilon^\ast<0$, and differentiating \eqref{eq:tn_rate} gives
\begin{align}
    \frac{\partial \bar{R}_{TN}}{\partial \varepsilon^\ast}
    &= \frac{W}{K}\Bigg[
    \frac{\partial M_{\mathrm{on}}}{\partial \varepsilon^\ast}
    \mathbb{E}[\log_2(1+\gamma_i)] \notag\\
    &\quad + M_{\mathrm{on}}
    \frac{\partial \mathbb{E}[\log_2(1+\gamma_i)]}
    {\partial \varepsilon^\ast}
    \Bigg]<0.
    \label{eq:derivative}
\end{align}
Thus, additional terrestrial energy savings through \ac{MBS} shutdown can come at the expense of lower per-\ac{UE} throughput due to increased loading and less favorable serving links.

\paragraph{Energy savings and trade-off.}

From~\eqref{eq:energy}, 
deactivating terrestrial \ac{MBS} $j$ reduces its power consumption by $p_j+\psi_j$.
Assuming a common power saving $\Delta P_{\mathrm{off}}$ per deactivated \ac{MBS}, 
the aggregate terrestrial power reduction is
\begin{equation}
    \Delta P_{TN}(\varepsilon^\ast) =
    \big(T-M_{\mathrm{on}}(\varepsilon^\ast)\big)
    \Delta P_{\mathrm{off}}.
    \label{eq:energy_saving}
\end{equation}
In the considered operating regime, 
increasing \ac{NT} offloading enables additional shutdowns,
such that $\partial M_{\mathrm{on}}/\partial\varepsilon^\ast<0$ and hence $\partial\Delta P_{TN}/\partial\varepsilon^\ast>0$.
Combining this with~\eqref{eq:derivative} yields
\begin{equation}
    \frac{d\bar{R}_{TN}}{d\Delta P_{TN}} =
    \frac{\partial\bar{R}_{TN}/\partial\varepsilon^\ast}
         {\partial\Delta P_{TN}/\partial\varepsilon^\ast}<0.
    \label{eq:pareto}
\end{equation}
This characterizes the expected trade-off between terrestrial per-\ac{UE} throughput and power savings under the considered operating conditions.

\section{Results and Discussion}
\label{sec:results}

We evaluate a terrestrial network of 1776 \acp{MBS} deployed over approximately \SI{2500}{km^2},
comprising an urban region with \SI{500}{m} inter-site distance and 40\% of the \ac{UE} population, 
and a rural region with \SI{1732}{m} inter-site distance and the remaining 60\%.
The \ac{UE} population follows the 24-hour traffic profile in Table~\ref{proportion_hour},
varying from approximately 400 \acp{UE} at 06:00 to 10{,}000 at 21:00, with 80\% of \acp{UE} located indoors.
Unless otherwise stated,
results are averaged over the complete 24-hour period, following~\cite{alam2023throughput}. Low-traffic results refer to hours 1--7, during which traffic is lowest according to Table~\ref{proportion_hour}.

The LEO satellite and the single HAPS (SC2) are located above the center of the study area. In SC3, the four HAPS are placed at $(\pm 16, \pm 16)$~km, while in SC4 the nine HAPS form a $3\times3$ grid with 22~km spacing. In SC3 and SC4, each HAPS can serve only UEs above a minimum elevation angle $\theta_{\min}$, corresponding to a footprint radius $r = h_{\mathrm{HAPS}}/\tan\theta_{\min}$ of 17.5~km ($\theta_{\min} \approx 48.8^\circ$) and 13.5~km ($\theta_{\min} \approx 56.0^\circ$), respectively, with neighboring footprints overlapping.

\subsubsection{UE association and bandwidth allocation}

\begin{figure}[b]
    \centering
    \includegraphics[width=0.7\linewidth]{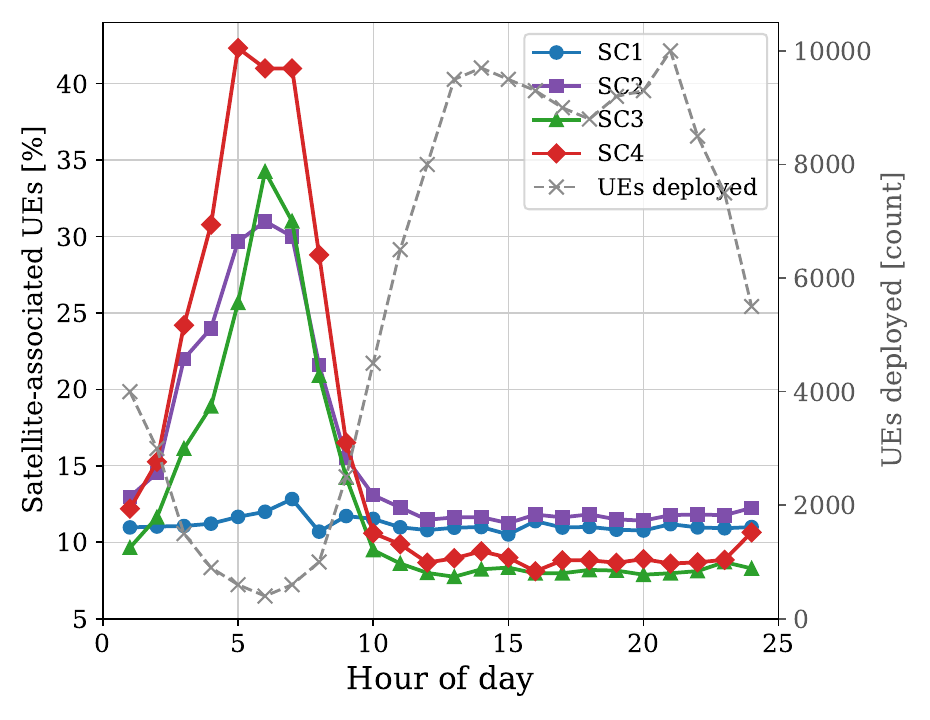}
    \caption{The fraction of \acp{UE} associated to the \ac{NT} tier.}
    \label{fig:assoc}
\end{figure}

Fig.~\ref{fig:assoc} shows the fraction of \acp{UE} associated with the \ac{NT} tier over the diurnal traffic profile.
BLASTER jointly optimizes \ac{UE} association, bandwidth allocation, and terrestrial \ac{MBS} activation to balance network performance and energy consumption.
For the multi-\ac{HAPS} configurations,
the \ac{NT} association fraction increases during low-traffic periods,
peaking around 06:00,
when offloading facilitates additional terrestrial \ac{MBS} shutdowns.
As traffic increases,
a larger fraction of \acp{UE} is served by the terrestrial tier.
Across configurations, 
the \ac{NT} association fraction follows SC4 $>$ SC3 $>$ SC2 $>$ SC1.
Compared with \ac{LEO},
\ac{HAPS} benefit from shorter propagation distances,
while multiple \ac{HAPS} provide more localized coverage and spatial frequency reuse, 
which BLASTER uses to increase offloading.


\subsubsection{Throughput performance}

\begin{figure}[!htb]
    \centering
    \includegraphics[width=.7\linewidth]{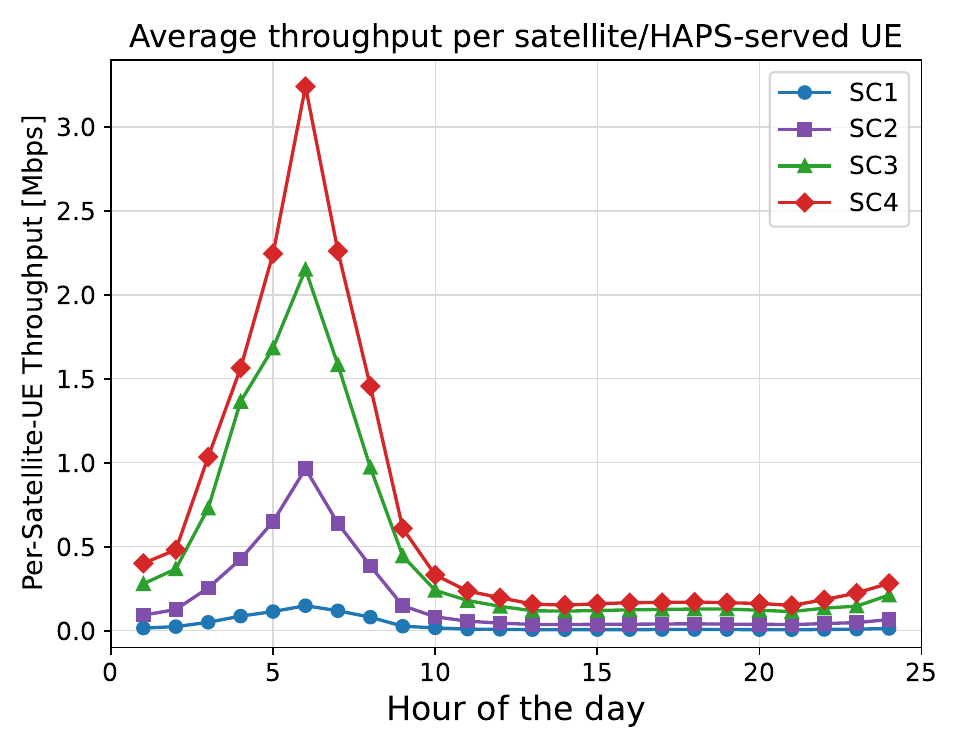}
    \caption{Average throughput per NT-served UE.}
    \label{fig:sat_thr}
\end{figure}

\begin{figure}[!htb]
    \centering
    \includegraphics[width=0.7\linewidth]{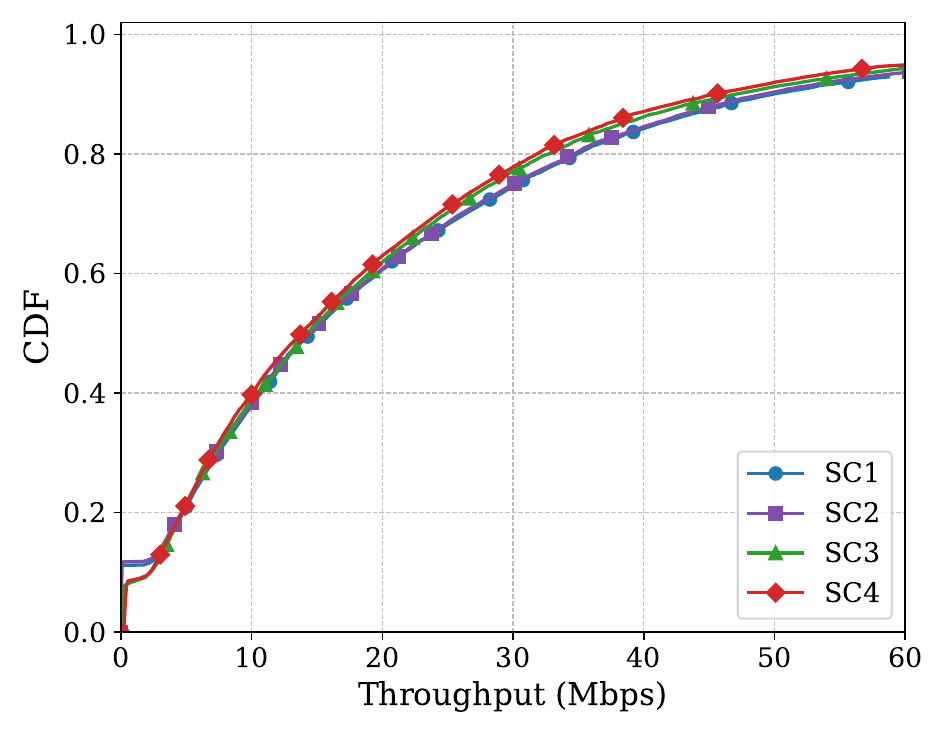}
    \caption{Per-UE throughput distribution in TN \& NTN layers.}
    \label{fig:TN_NTN throughput}
\end{figure}

Fig.~\ref{fig:sat_thr} shows the average throughput per \ac{NT}-served \ac{UE},
reflecting the combined effects of propagation conditions, spatial frequency reuse, and BLASTER's association and bandwidth-allocation decisions.
Increasing the number of \ac{HAPS} improves this throughput:
averaged over 24~hours, SC4 achieves 38\% and 268\% higher values than SC3 and SC2, respectively,
and approximately $20\times$ higher than \ac{LEO} (SC1).
These gains remain nearly unchanged during the low-traffic hours 1--7 (38\%, 255\%, and $20\times$, respectively),
and the hourly SC4-to-SC1 ratio remains between $18\times$ and $24\times$ over the entire day.
This large gain results from two multiplicative effects.
First, the bandwidth available per \ac{NT}-served \ac{UE} in SC4 is approximately $11\times$ that of SC1,
owing to the larger \ac{NT} bandwidth fraction $\varepsilon$ and to the reuse of the \ac{NT} band across the nine \ac{HAPS}.
Second, the shorter propagation distance of the \ac{HAPS} links increases the average spectral efficiency by approximately $1.6\times$.
Fig.~\ref{fig:TN_NTN throughput} shows the whole-network per-\ac{UE} throughput distribution, 
with SC4 performing slightly worse than SC1 and SC2 despite its higher \ac{NT}-tier throughput; the mean per-\ac{UE} throughput is 17.3~Mbps for SC4, compared with 19.2~Mbps for SC1.
This behavior is consistent with the trade-off discussed in~\eqref{eq:derivative},
as additional terrestrial \ac{MBS} shutdowns can increase the load on remaining cells and force reassociated \acp{UE} onto more distant serving \acp{MBS}.
The resulting reduction in available bandwidth per \ac{UE} and received signal power may outweigh the interference reduction from shutdown,
lowering terrestrial throughput.
Thus, the larger terrestrial power savings characterized by~\eqref{eq:energy_saving} can come at the expense of a modest degradation in whole-network throughput performance.

\subsubsection{Terrestrial MBS shutdown and energy savings}

\begin{figure}[t]
\centering
\includegraphics[width=0.7\linewidth]{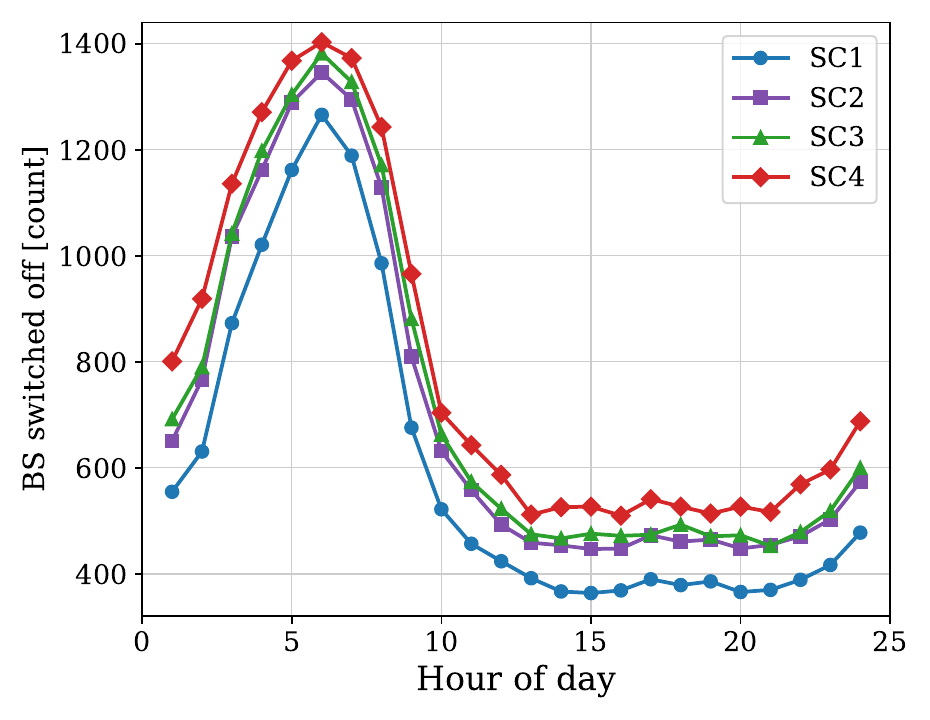}
\caption{Daily MBS shut-down}
\label{fig:MBS_shut}
\end{figure}

\begin{figure}[t]
\centering\includegraphics[width=0.7\linewidth]{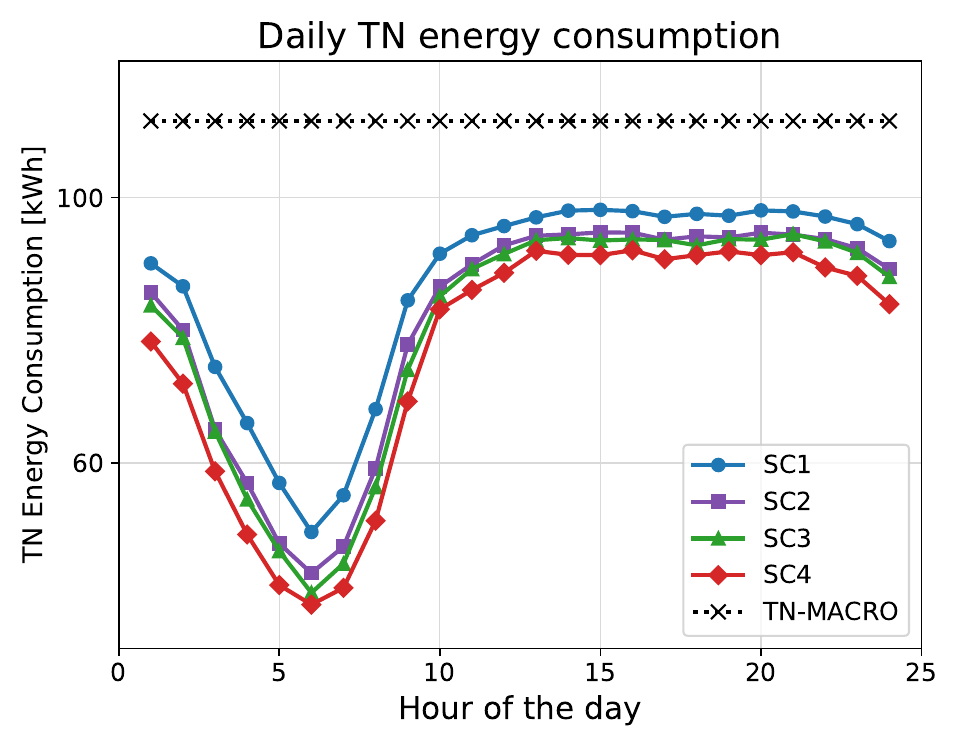}
\caption{TN vs NTN energy consumption.}
\label{fig:energy}
\end{figure}

Fig.~\ref{fig:MBS_shut} shows the number of terrestrial \acp{MBS} deactivated over the diurnal traffic profile.
BLASTER exploits \ac{NT} offloading to deactivate lightly loaded terrestrial \acp{MBS} while maintaining \ac{UE} coverage.
The shutdown count peaks around 06:00, 
when traffic demand is low, 
and decreases as traffic increases.
Across the considered configurations,
the number of deactivated \acp{MBS} follows SC4 $>$ SC3 $>$ SC2 $>$ SC1, 
reflecting the greater offloading opportunities provided by multi-\ac{HAPS} deployments.
As indicated by~\eqref{eq:energy_saving},
additional \ac{MBS} shutdowns translate into greater terrestrial power savings.
Fig.~\ref{fig:energy} shows the corresponding terrestrial energy consumption,
while Table~\ref{tab:energy_windows} summarizes the energy reduction relative to the \ac{TN}-MACRO baseline.
Over the full day,
terrestrial energy consumption decreases by 26.0\%, 30.3\%, 31.3\%, and 34.1\% for SC1, SC2, SC3, and SC4, respectively.
The reductions increase during low-traffic hours,
reaching 54.7\%, 58.1\%, 59.7\%, and 60.6\%, respectively, at the quietest hour.
Compared with SC1, SC4 achieves an additional 8.1 percentage points of daily energy savings,
increasing to 9.7 percentage points during the low-traffic period.
These results demonstrate how BLASTER exploits the greater offloading capabilities of multi-\ac{HAPS} deployments to reduce terrestrial layer energy consumption.

\begin{table}[h]
\centering
\caption{TN energy reduction relative to the TN-MACRO.}
\label{tab:energy_windows}
\begin{tabular}{lccc}
\toprule
Tier & 24h & Low-traffic & Quiet-hr \\
\midrule
SC1    & 26.0\% & 41.3\% & 54.7\% \\
SC2  & 30.3\% & 46.5\% & 58.1\% \\
SC3  & 31.3\% & 47.7\% & 59.7\% \\
SC4  & 34.1\% & 51.0\% & 60.6\% \\
\bottomrule
\end{tabular}
\end{table}

\subsubsection{QoS-energy trade-off}
\begin{figure}[t]
    \centering
    \includegraphics[width=0.8\linewidth]{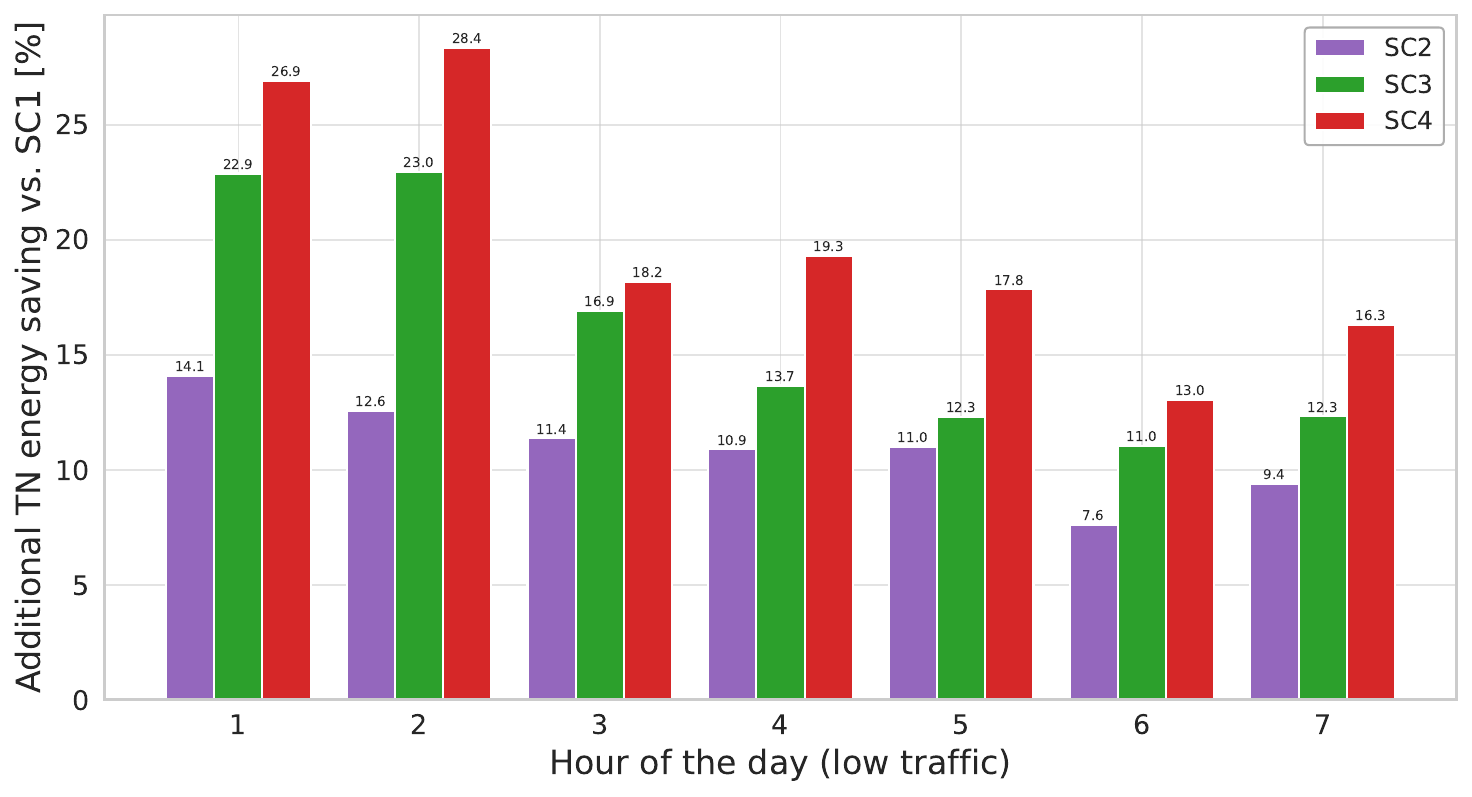}
    \caption{Additional terrestrial energy saving of SC2--SC4 relative to SC1 at the SLT of SC1.}
    \label{fig:figT1_1}
\end{figure}

\begin{figure}[t]
    \centering
    \includegraphics[width=0.8\linewidth]{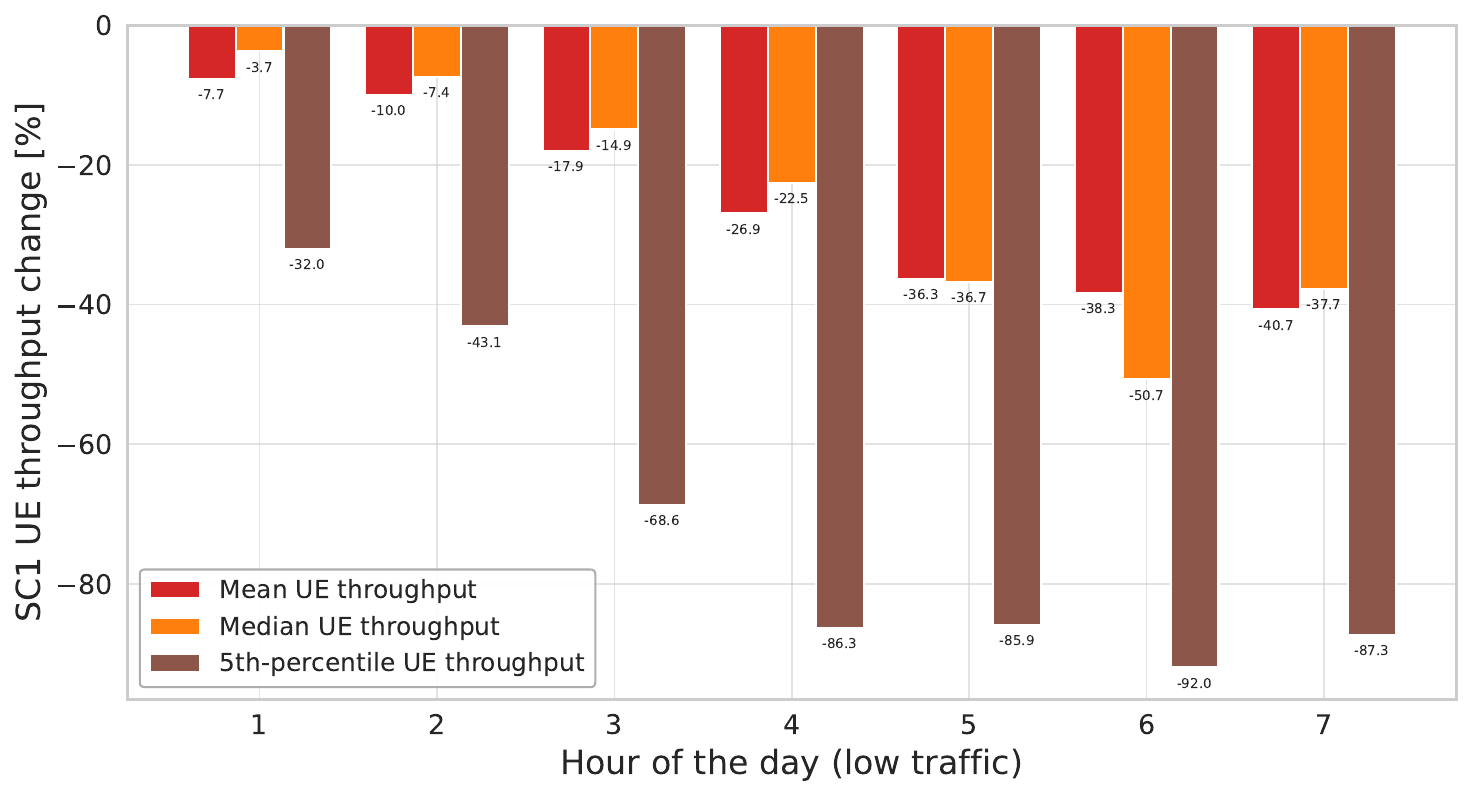}
    \caption{UE throughput change of SC1 when forced to deactivate as many terrestrial MBSs as SC4.}
    \label{fig:figT2_2}
\end{figure}

To quantify the trade-off characterized in~\eqref{eq:pareto}, we use the SLT of SC1 in each hour as the reference \ac{QoS}. Starting from the BLASTER solution, additional terrestrial \acp{MBS} are switched off greedily, least-loaded first, and their \acp{UE} are reassociated with the active \ac{MBS} providing the highest RSRP. For Fig.~\ref{fig:figT1_1}, this procedure is applied to SC2--SC4 until their SLT reaches that of SC1; for Fig.~\ref{fig:figT2_2}, it is applied to SC1 until it deactivates as many \acp{MBS} as SC4. Both analyses cover the low-traffic hours 1--7.
As shown in Fig.~\ref{fig:figT1_1}, at the same SLT as SC1, SC2, SC3, and SC4 reduce the terrestrial energy consumption by an additional 11.0\%, 16.0\%, and 20.0\% on average, respectively, with SC4 reaching 28.4\%. The gain is largest in the first hours, when more traffic can be offloaded to the \ac{HAPS} tier.
Conversely, Fig.~\ref{fig:figT2_2} shows that forcing SC1 to deactivate as many \acp{MBS} as SC4 reduces its mean, median, and 5th-percentile \ac{UE} throughput by 25.4\%, 24.8\%, and 70.7\% on average, respectively, as reassociated \acp{UE} are pushed to distant and more loaded \acp{MBS}. At the same terrestrial energy consumption, SC4 achieves a mean \ac{UE} throughput comparable to that of SC1 (69.2 vs.\ 69.6~Mbps), while raising the 5th-percentile throughput from about 0.01 to 4.9~Mbps, at the cost of a lower median throughput. Hence, multi-\ac{HAPS} deployments enable deeper TN energy savings while substantially protecting cell-edge \acp{UE}.

\section{Conclusion}
We extended BLASTER to multi-HAPS deployments to investigate the trade-off between network QoS and terrestrial energy consumption. By jointly optimizing UE association, NT bandwidth allocation, and terrestrial transmit power, with a sparsity-promoting relaxation that drives lightly loaded MBSs to shutdown, the framework adapts terrestrial activation to the diurnal traffic profile. The 9-HAPS configuration reduces terrestrial energy consumption by 34.1\% over the full day and 60.6\% at the quietest hour, compared with 26.0\% and 54.7\% for LEO. At the same SLT, the 9-HAPS configuration achieves 20\% additional terrestrial energy savings relative to LEO, while substantially improving the throughput of cell-edge UEs, demonstrating how enhanced NT offloading can reduce terrestrial infrastructure energy consumption. Future work will investigate heterogeneous HAPS and joint LEO-HAPS deployments, and extend the joint optimization to NT transmit power and HAPS placement.

\bibliographystyle{IEEEtran}
\bibliography{bibliography}
\end{document}